\documentclass[a4paper]{article}
\pdfoutput=1

\usepackage{amssymb}
\usepackage{arydshln}
\usepackage{color}
\usepackage{graphicx}

\begin{document}

\begin{flushleft}

\begin{Large}
Visibility graph analysis of wall turbulence time-series\\
\end{Large}
\vspace{0.5cm}
Giovanni Iacobello$^{1}$\footnote[1]{Corresponding Author. giovanni.iacobello@polito.it\\ stefania.scarsoglio@polito.it; luca.ridolfi@polito.it}, Stefania Scarsoglio$^1$, Luca Ridolfi$^2$\\
\vspace{0.5cm}
\begin{small}
$^1$ Department of Mechanical and Aerospace Engineering, Politecnico di Torino, 10129 Torino, Italy\\
$^2$ Department of Environmental, Land and Infrastructure Engineering, Politecnico di Torino, 10129 Torino, Italy
\end{small}
\end{flushleft}

\begin{abstract}

	 The spatio-temporal features of the velocity field of a fully-developed turbulent channel flow are investigated through the natural visibility graph (NVG) method, which is able to fully map the intrinsic structure of the time-series into complex networks. Time-series of the three velocity components, $(u,v,w)$, are analyzed at fixed grid-points of the whole three-dimensional domain. Each time-series was mapped into a network by means of the NVG algorithm, so that each network corresponds to a grid-point of the simulation. The degree centrality, the transitivity and the here proposed mean link-length were evaluated as indicators of the global visibility, inter-visibility, and mean temporal distance among nodes, respectively. The metrics were averaged along the directions of homogeneity ($x, z$) of the flow, so they only depend on the wall-normal coordinate, $y^+$. The visibility-based networks, inheriting the flow field features, unveil key temporal properties of the turbulent time-series and their changes moving along $y^+$. Although intrinsically simple to be implemented, the visibility graph-based approach offers a promising and effective support to the classical methods for accurate time-series analyses of inhomogeneous turbulent flows.
	
\end{abstract}

\begin{flushleft}
Keywords: Turbulent channel flows; Complex networks; Time-series analysis; Visibility graph; Direct numerical simulations.
\end{flushleft}

\section{Introduction}

		One of the most challenging research topic in classical physics is represented by turbulent flows. Their great importance is evident through a number of natural phenomena (e.g., rivers, atmospheric and oceanic streams), industrial and civil applications (e.g., flow through pumps, heat exchangers, wake flows of vehicles and aircraft, wind-building interactions) in which turbulence is involved. The study of wall-bounded turbulent flows, in particular, is a very active research field, due to the great attention paid to the fluid-structure interaction. Although deeply studied from a phenomenological and theoretical point of view, the turbulence dynamics, due to their complexity, are still not fully understood \cite{warhaft2002,Smits2011}. Nowadays, several numerical simulations and experiments are performed, providing a massive amount of spatio-temporal data that needs to be properly examined. Different approaches, mainly relying on statistical techniques, are then typically used to explore and analyze turbulent flows.
		
		Among all the proposed techniques, time-series analysis is a broadly adopted approach to study the temporal evolution of dynamical systems, specifically those with high intrinsic complexity. Different methods, such as Fourier and wavelet transforms \cite{box2015,percival2006}, as well as nonlinear approaches \cite{strogatz1994,kantz2004,Campanharo2011}, have been developed so far to extract information from time-series. However, since each method unavoidably loses some information about the temporal structure of the series analyzed, new approaches are continuously required to fill this lack. In the last decades, complex networks, by combining elements from graph theory and statistical physics \cite{WattsStrogatz1998,Boccaletti2006,Newman2003}, have turned out to be powerful tools to study complex systems, specifically by mapping time-series to extract non-trivial information \cite{gao_epl_2016}. Recently, several improvements were gained in this field and numerous advances were proposed based on different approaches \cite{nunez2012}, such as correlation \cite{ZhangSmall2006,YangYang2008}, visibility \cite{Lacasa2008,Luque2009}, phase-space reconstruction \cite{Xu2008,Gao2009}, recurrence quantification \cite{Donner2010,gao2013,gao_chaos_2017}, and transition probabilities \cite{Nicolis2005,Shirazi2009} algorithms.

\noindent Beside the well-established applications to Internet, World Wide Web, economy and social dynamics \cite{Costa2007,Havlin}, growing attention has been given nowadays to the application of complex networks to fluid flows and different flow regimes have been explored, such as two-phase flows \cite{Gao2009,gao2013,gao_ijbc_2017}, geophysical flows \cite{lindner2017,tupikina2016}, turbulent jets \cite{Shirazi2009,charakopoulos2014}, reacting flows \cite{murugesan2015}, as well as fully developed turbulent flows \cite{liu2010,Manshour2015_fullydev} and isotropic turbulence \cite{taira2016, Scarsoglio2016}.
		
		In this work, the natural visibility algorithm is exploited to investigate the spatio-temporal characterization of a fully-developed turbulent channel flow, solved through a direct numerical simulation (DNS) and available from the Johns Hopkins Turbulence Database (JHTDB) \cite{JHTD1,JHTD2}. Time-series of the three velocity components were analyzed at fixed spatial positions, and a single network was built at each point. In so doing, an ensemble of networks was obtained, where each network corresponds to a time-series. This novel approach allows us to capture some important aspects of the temporal structure of the signal and how these features change along the wall-normal direction. In fact, we can systematically extract information about the occurrence and temporal collocation of extreme events (i.e., peaks) and irregularities, which are fundamental features to characterize turbulent flows. The statistical tools classically adopted in turbulence, such as correlation function, higher-order statistics, structure functions, energy spectrum and probability density functions, all fail in preserving and discerning the temporal structure of a time-series (e.g., two different temporal signals can have the same probability density function or energy spectrum). The visibility approach here presented is instead able to fully inherit and point out the temporal structure of the turbulent series: the different temporal dislocation of events such peaks and fluctuations will lead, case by case, to a different network topology.

\noindent A systematic approach to highlight temporal features of the time-series through the most significant network metrics is thus proposed and discussed. Particular care is given not only to relate the network topology to the temporal structure of the series, but also paying attention to the physical interpretation of the results. New insights into how the network topology is affected by important temporal features of the mapped signal are thus provided. Specific combinations of the trend of the network metrics are able to shed light into the time-series structure. Furthermore, a qualitative correspondence between the network metrics and the flow dynamics is presented, underlying the ability of the method to identify different flow regions.
		
\section{Methods}\label{sec:methods}
	\subsection{Database description}\label{sec:db}

		The data here used were extracted from a DNS of a fully developed turbulent channel flow \cite{JHTD3}, available from the JHTDB \cite{JHTD1,JHTD2}. The simulation is performed at $Re_\tau=1000$, where $Re_{\tau}=h u_{\tau}/\nu$ is the friction velocity Reynolds number, $h=1$ is the half-channel height, $\nu = 5 \cdot 10^{-5}\, U_b h$ is the viscosity, $U_b=1$ is the bulk channel velocity, and $u_{\tau}=5 \cdot 10^{-2}$ is the friction velocity (all physical parameters are dimensionless). Periodic boundary conditions in the streamwise ($x$) and spanwise ($z$) directions are adopted, while the no-slip condition is imposed at the top and bottom walls, $y/h=\pm 1$ ($y$ is the direction normal to the wall). Once the statistically stationary conditions were reached, the simulation was carried on for approximately one flow-through time, $t \in [0,26]h/U_b$, with a storage temporal step $\delta t =0.0065$. Thus $N_t=4000$ temporal frames are available. Velocity $(u, v, w)$ and pressure $(p)$ fields were computed over the physical domain, $(L_x \times L_y \times L_z) = (8\pi h \times 2h \times 3\pi h)$, and stored with a grid resolution $(N_x \times N_y \times N_z) = (2048 \times 512 \times 1536)$. Other simulation parameters and flow statistics are given elsewhere \cite{JHTD3}.

In this study, a subset of the domain was taken into account, exploiting the geometrical features of the flow field along the three directions $(x,y,z)$. In the wall-normal direction, $y$, due to the geometric symmetry, only grid-points from the bottom wall to the half-channel height, $-1\leq y/h \leq 0$, were considered. As a result, the values of the dimensionless distance from the wall, $y^+$, defined as \cite{pope} $y^+= (y/h+1) Re_\tau$, ranges in the interval $[0, 10^3]$. Along the $y$-direction the distance between consecutive grid-points was selected to increase gradually from the wall towards the center of the channel. Indeed, close to the wall the flow is strongly inhomogeneous, and a finer spatial resolution is necessary to better capture the features of the flow field. Differently, along the $x$ and $z$ directions, a coarse uniform storage was adopted. In fact, along these two directions the flow is statistically homogeneous and fewer uniformly spaced grid-points are sufficient to guarantee the statistical stationarity of the results.

\noindent The selected sub-domain size is $(S_X \times S_Y \times S_Z)=(64  \times 70  \times 12)$, where the first grid-point $Y=0$ corresponds to the wall coordinates $y/h=-1$ and $y^+=0$. Details of the sub-domain structure are reported in Appendix A.


	\subsection{Mapping time-series into networks: the visibility algorithm}\label{sec:visi}

In the time-series analysis, complex networks represent a recent and promising tool to highlight and characterize important structural properties\cite{Campanharo2011, Boccaletti2006}. In the present work, the natural visibility algorithm proposed by Lacasa and co-authors \cite{Lacasa2008} was adopted. According to this method, two values $(t_i, s(t_i))$ and $(t_j, s(t_j))$ of a univariate time-series $s(t_n)$, $n=\lbrace 1,2,...,N\rbrace$, have visibility, and consequently are two connected nodes of the associated network, if the following condition
		\begin{equation}\label{eq:visibility}
			s(t_k)<s(t_j)+\left(s(t_i)-s(t_j)\right)\frac{t_j-t_k}{t_j-t_i},
		\end{equation}
		is fulfilled for any $t_i<t_k<t_j$ (or equivalently $i<k<j$). From a geometrical point of view, two nodes are linked if there is a straight line connecting them without intersecting any intermediate data. The natural visibility criterion is, therefore, a \textit{convexity criterion}. A geometrically simpler version of the NVG can be obtained considering only horizontal lines among data, defining the so called horizontal visibility graph \cite{Luque2009}. In this case, the horizontal visibility satisfies an ordering criterion.
		
The visibility algorithm is simple to implement and has been applied in many different fields (e.g., \cite{stephen2015,supriya2016,donner2012,mutua2016,gao_neur_syst_2017}), including fluid flows \cite{liu2010,charakopoulos2014,murugesan2015,Manshour2015_fullydev,Singh2017,braga2016}. However, the visibility approach has some drawbacks, related to the fact that it is invariant under affine transformations \cite{Lacasa2008} (i.e., rescaling and translation of both horizontal and vertical axes), so this could lead to a lost of information in mapping the time-series. Moreover, if time-series with a considerable number of observations (indicatively $N_t>10^4$) are analyzed, then the condition (\ref{eq:visibility}) requires to be verified many times. In these situations, as in this study, an optimized approach is crucial to sharply decrease the computational costs (e.g., see \cite{Lan2015}).

	\subsection{Complex network metrics}\label{sec:CN_metr}
A summary of the network metrics investigated in the present work is here reported \cite{Boccaletti2006,Costa2007,Newman2003}. A network is defined as a graph $G(N,E) = (\mathcal{V},\mathcal{E})$, where $\mathcal{V}=\lbrace 1,2,...,N\rbrace$ is a set of $N$ labeled nodes (or vertices) and $\mathcal{E}=\lbrace 1,2,...,E\rbrace$ is a set of $E$ links (or edges), with non-trivial topological features. The \textit{adjacency matrix}, $A_{ij}$, defined as

		\begin{equation} \label{eq:AdjM}
		A_{ij} = \left \{\begin{array}{l} 0$, if $\lbrace i,j\rbrace\not\in\mathcal{E},\\ 1$, if $\lbrace i,j\rbrace\in\mathcal{E}, 	
		\end{array}	
		\right.
		\end{equation}

\noindent determines the existence of a link between a pair $\lbrace i,j\rbrace$ of nodes. Beside being unweighted (A is a binary matrix), in this study we only consider undirected networks ($A_{ij}=A_{ji}$) with no self-loops ($A_{ii}=0$).

		In general, two kinds of metrics can be defined: metrics associated to nodes, namely \textit{local metrics}, and metrics related to the entire network, here referred as \textit{global metrics}. In the following, the $\langle \bullet \rangle$ notation is adopted to indicate that global metrics were obtained by averaging (over all nodes in the network) the corresponding local metrics.
		
		The \textit{degree centrality} of a node is defined as
		\begin{equation} \label{eq:DC}
			k_i=\sum_{j=1}^N{A_{ij}},
		\end{equation}
		and gives the number of topological neighbors of node $i$, that is the number of nodes linked to it (the set of neighbors is called the \textit{neighborhood}, $\Gamma_i$, of node $i$). In particular, if all $N$ nodes in a network are linked each other, the network is said \textit{fully-connected} and the degrees are all constant and equal to $N-1$. The average degree centrality of a network is then
		\begin{equation} \label{eq:DC_m}
			\langle k\rangle=\frac{1}{N}\sum_{i=1}^N{k_i}.
		\end{equation}
		As a centrality measure, the degree, $k_i$, is an indicator of the most important vertices in a network. The fraction of nodes in the network that have degree $k$ is the \textit{degree distribution}, $p_k$, and it also represents the probability that a randomly chosen node has degree $k$. In many real networks, $p_k$ is heavy-tailed, because of an intrinsic noise due to the finiteness of the time-series \cite{Boccaletti2006}. In these cases it may be useful to evaluate the \textit{cumulative degree distribution} \cite{Newman2003}:
		\begin{equation} \label{eq:Pkcum}
			P_k=\sum_{k'=k}^{\infty}{p_{k'}},
		\end{equation}		
		which is the probability to find a degree greater than or equal to $k$. The statistical fluctuations present in the tails of the $p_k$ distribution are smoothed if $P_k$ is used \cite{Boccaletti2006}.
			
The \textit{transitivity}, $Tr$, is a global clustering metric and is defined as \cite{Costa2007}
		\begin{equation} \label{eq:Trans}
			Tr=\frac{3N_\Delta}{N_3},
		\end{equation}
		where $N_\Delta$ and $N_3$ are the number of triangles and the number of
connected triples in the network, respectively. A triangle is a set of three nodes linked between them. A connected triple, instead, is a set of three nodes where two of them must be directly linked to the third node. The transitivity, $0 \leq Tr \leq 1$, is therefore a measure of the presence of triangles in the network. Another commonly used clustering metric is the \textit{clustering coefficient}, $C_i$ \cite{Boccaletti2006}. Both the transitivity and the clustering coefficient are measures of the presence of triangles in the network, but $C_i$ tends to weight the contributions of low-$k_i$ vertices more heavily that $Tr$ \cite{Newman2003}, being its denominator proportional to $k_i^2$, and making its interpretation less clear and general. For this reason, in the following we only focus on the transitivity to capture the inter-node relations among nodes.

		Finally, we propose a new local metric, based on the temporal length between two mutually visible vertices \cite{Bezsudnov2014}, defined as the \textit{mean link-length}:
		\begin{equation} \label{eq:d1n_i}
			d_{1n}(i)=\frac{1}{k_i} \sum_{j\in\Gamma_i} |t_j-t_i|,	
		\end{equation}
		
\noindent where $\Gamma_i$ and $k_i$ are the neighborhood and the degree centrality of node $i$, respectively. From a time-series point of view, each node represents a temporal event and, then, the physical distance between two nodes $i$ and $j$ can be defined as $|t_j-t_i|$. It follows that $d_{1n}$ increases when a node is linked to nodes far in time from it. Averaging over all nodes in the network, a global measure is obtained:
		\begin{equation} \label{eq:d1n_m}
			\langle d_{1n}\rangle=\frac{1}{N} \sum_{i=1}^N {d_{1n}(i)}.	
		\end{equation}
		
	\subsection{Building the networks} \label{sec:building}

In this study, the velocity field $(u,v,w)$ was focused on, being one of the most basic and significant field to analyze a turbulent flow. Exploiting the visibility-invariance under affine transformations, in the following each time-series is normalized as $u^*(t_i)=(u(t_i)-\mu)/\sigma$, where $\mu$ and $\sigma$ are the local mean and standard deviation values of $u(t_i)$, respectively. By following the classical decomposition adopted for the statistical description of turbulence \cite{pope}, the resulting signal, $u^*$, has zero mean value and standard deviation equal to 1. At fixed point, $(u^*,v^*,w^*)$ represent the net turbulent fluctuations of the velocity field. The adopted decomposition allows one to separate the complete signal, $u(t)$, into a mean term constant in time, $\mu$, and a fluctuating temporal part, $u^*(t)$. In so doing, we can focus on the temporal variations only, by comparing normalized signals having the same mean and standard deviation values. Turbulent fluctuations are the basis of the statistical description of turbulence. For example, root-mean-square velocity, $u_{rms} = \sqrt{\overline{u^*(t)^2}}$ (the overbar represents the temporal average), is usually defined to quantify the turbulence strength. A high $u_{rms}$ indicates an elevate level turbulence. Thus, $(u^*,v^*,w^*)$ hold the primary indication of the turbulence intensity of a velocity field.

\noindent For each grid-point in space, all the $N_t=4000$ time frames were then exploited to build the networks, being the velocity series dependent only on the time (i.e., the series are univariate at fixed coordinates). Therefore, $S_x \times S_y \times S_z = 53760$ networks were constructed for each velocity component. According to the visibility algorithm --- since each of the resulting networks is connected (i.e. every node has at least one neighbor) \cite{Lacasa2008} --- each temporal instant corresponds to a node. Consequently, all the $53760$ networks have the same number of nodes $N=N_t=4000$. The number of links of each network, $E$, instead, can be obtained from the average degree values by applying the general relation, $E=\langle k\rangle N/2$. A sensitivity analysis on the number of nodes $N$ is reported in Appendix B, for which different temporal discretizations (namely $2 \, \delta t$ and $4 \, \delta t$, where $\delta t=0.0065$ is the temporal discretization leading to 4000 time frames) are considered, resulting in different cardinality of the networks.
		
		We recall that in a fully developed turbulent channel flow the velocity and pressure fields are statistically homogeneous along the streamwise, $x$, and spanwise, $z$, directions. The wall normal coordinate, $y^+$, is then the only direction where spatial inhomogeneities develop. Since network measures inherit the properties of the mapped time-series \cite{Lacasa2008}, also the global metrics (i.e. averaged over the nodes of each network) were assumed as statistically homogeneous in the $x$-$z$ directions (more details are reported in Appendix C, where few representative plots of the global metrics calculated at fixed $x$ and $z$ locations are reported). Consequently, the global metrics were firstly calculated for each single network. Subsequently, such global metrics were averaged over the $S_x \times S_z$ grid-points as:
		\begin{equation} \label{eq:aver_measures}
			\widetilde{\mathcal{F}}(Y) =\frac{1}{S_xS_z}\sum_X\sum_Z{\mathcal{F}(X,Y,Z)},
		\end{equation}
		where where $\mathcal{F}$ represents the specific metric considered, namely $\mathcal{F}=\left\lbrace \langle k \rangle; Tr; \langle d_{1n}\rangle\right\rbrace$. In so doing, we obtained three averaged quantities: $\widetilde{k}$, $\widetilde{Tr}$, and $\widetilde{d_{1n}}$, where the notation $\widetilde{(\bullet)}$ indicates the average over grid-points in the directions of spatial homogeneity of the flow. This operation makes the above averaged metrics dependent only on the wall normal coordinate $y^+$ and their plots statistically meaningful.
		
\section{Relating time-series structure and network metrics} \label{sec:relations}
			It is known that the general structure of time-series is preserved in the topology of the associated natural visibility graphs, as shown by Lacasa et al. \cite{Lacasa2008} and as emerges from successive works \cite{charakopoulos2014,stephen2015,murugesan2015,Manshour2015_fullydev,liu2010,donner2012}. Specifically, periodic time-series are converted into \textit{regular networks}, i.e. graphs where nodes have constant degrees related to the periods of the series. Fractal series, instead, convert into networks with power-law degree distributions \cite{lacasa2009estim}. In particular, fractional Gaussian noise with Hurst exponent equal to $0.5$ (i.e., uncorrelated random series) are mapped by the NVG method into networks with power-law degree distribution with exponent, $\gamma$, equal to $4$ \cite{lacasa2009estim}.
			
\noindent Here, particular attention is paid relating the network topology and the temporal structure of the series to a physical interpretation of the network metrics, with respect to the flow dynamics. In fact, although the overall features of the time-series are inherited by the corresponding visibility graphs, it is not straightforward how topological network metrics are affected by different temporal behaviors of the series. In order to explore this gap, we qualitatively relate the global metrics behavior to the temporal structure of the corresponding time-series.
			
			In general, if two different time-series are compared, they can differ in several ways. In this analysis, we focused on the presence of \textit{peaks} and \textit{irregularities}. A point of a time-series, $s(t_i)$, is said a \textit{peak} if it is a local (or global) maximum of $s(t_i)$, with order of magnitude comparable with the maximum excursion of the series, $\Delta= s_{max}-s_{min}$. Peaks generally have higher probabilities to connect to other points in the series, because obstacles to the visibility are avoided from higher positions. However, in turn, the long-term visibility of points in the surroundings of peaks is obstructed by the peaks, thus creating local barriers to the visibility of lower points of the series. \textit{Irregularities} are temporal variations with order of magnitude much smaller than $\Delta$, and defined as local barriers decreasing the visibility of the surrounding points. Peaks and irregularities are focused mainly for two reasons. First, the occurrence and temporal collocation of extreme events (i.e., peaks) and irregularities represent some of the fundamental features to characterize turbulent flows. Second, the NVG is a suitable method to evidence this kind of flow properties and translate them into the network topology. In particular, among all the topological parameters investigated, the transitivity, $Tr$, the global mean link-length, $\langle d_{1n}\rangle$, and the average degree, $\langle k \rangle$, turned out to be the metrics that better capture the temporal structure of the time-series in terms of peaks and irregularities, inheriting important features of the turbulent flow dynamics.
		
		In order to schematize how the occurrence of peaks and irregularities affects the temporal structure and in turn the network topology, we consider four exemplifying time series, as reported in Fig. \ref{fig:example_dc_series}. The starting series (panel a) is a sine function. With respect to panel (a), in panels (b)-(d) a uniform random noise is added to account for irregularity, while in panels (c) and (d) the periodicity is halved. The graphical representation of the networks corresponding to each time-series is reported on the right panels of Fig. \ref{fig:example_dc_series}.
	
	\subsection{Transitivity analysis}
		Let us first focus on the transitivity, $Tr$. We recall that, since each pair $(j,l)\in\Gamma_i$ ($\Gamma_i$ is the neighborhood of node $i$) always forms a connected triple with node $i$, the total number of triples in the network depends on the size of all the neighborhoods $\Gamma$. On the other hand, triangles are formed only if the nodes $(j,l)\in\Gamma_i$ are also linked, that is if $A_{ij}=A_{il}=A_{jl}=1$. In general, short-term connections are the most probable ones because time-series are not expected to sharply change in time (except for random series), so that nodes which are close in time are more likely to form triangles. If two neighbors $(j,l)\in\Gamma_i$ are far in time, instead, there are many nodes in between $j$ and $l$ so that there is a high probability to find a node that obstructs the inter-visibility of $j$ and $l$. As a result, the total number of triangles and triples, and therefore the transitivity, strongly depend on the inter-visibility of nodes inside each neighborhood. The transitivity can be then interpreted as a measure to characterize the typical convexity properties on some intermediate time-scale (i.e., the neighborhood temporal lengths) \cite{donner2012}.
	
		To better describe the effects of peaks and irregularities on the transitivity, let us consider the time-series, $s(t_i)$, and the corresponding networks, $G$, reported in Fig. \ref{fig:example_dc_series}. Time-series $s(t_i)_{(a),(c)}$ are clearly more regular than the series $s(t_i)_{(b),(d)}$, while $s(t_i)_{(c)}$ and $s(t_i)_{(d)}$ have three peaks instead of two. Therefore, while the networks $G_{(a)}$ and $G_{(c)}$ are well organized in clusters (one cluster for $G_{(a)}$, two for $G_{(c)}$), $G_{(b)}$ and $G_{(d)}$ appear more complex (right panels of Fig. \ref{fig:example_dc_series}). This happens because in $G_{(b),(d)}$ there are many nodes with low visibility due to the presence of irregularities.
		
		More in detail, any point in the ranges $t_i=(1-50)$ and $t_i=(51-100)$ of $s(t_i)_{(c)}$ has basically the same inter-visibility of corresponding points in $s(t_i)_{(a)}$ (i.e., nodes at the same relative altitude), because there are no substantial local changes of regularity between $s(t_i)_{(a)}$ and $s(t_i)_{(c)}$. As a consequence, the value of transitivity of $G_{(a)}$ and $G_{(c)}$ are expected to be scarcely affected by different occurrence of peaks, as evident from the Fig. \ref{fig:example_metr} where $Tr_{(c)}$ and $Tr_{(a)}$ are actually almost the same. The presence of more (or less) peaks in a time-series then does not significantly modify the inter-visibility (i.e., the transitivity) of nodes. Therefore, also $Tr$ of $G_{(b)}$ and $G_{(d)}$ are almost equal (see Fig. \ref{fig:example_metr}), being the irregularities of time-series $s(t_i)_{(b)}$ and $s(t_i)_{(d)}$ very similar.
		
		On the other hand, the time-series $s(t_i)_{(b),(d)}$ clearly display irregularities if compared with time-series $s(t_i)_{(a),(c)}$. The inter-visibility among neighbors of a generic node is obstructed because of the irregularities in the time-series. Let us consider an arbitrary node, for example $i=34$, highlighted as a green-colored dot in panels (a) and (b) of Fig. \ref{fig:example_dc_series}. While in $G_{(a)}$ the neighborhood $\Gamma_{34, (a)}$ (highlighted in yellow in Fig. \ref{fig:example_dc_series}) includes either short-term, medium-term, and long-term links, in $G_{(b)}$ the neighborhood $\Gamma_{34, (b)}$ includes only short-term and long-term connections. Therefore, the number of triangles (relative to the number of triples) in which is involved a generic node (e.g., $i=34$) is generally lower in irregular networks than in regular ones. As a result, the values of $Tr_{(b)}$ and $Tr_{(d)}$ are much lower than $Tr_{(a)}$ and $Tr_{(c)}$, as observed in Fig. \ref{fig:example_metr}. Summarizing, the transitivity is much more affected by local variations due to the presence of irregularities rather than the presence of local peaks in the series. In terms of flow dynamics, the transitivity is related to the presence of local fluctuations between consecutive peaks. Recalling that the all signals are normalized with respect to their mean and standard deviation values, the transitivity is thus a net measure of the intrinsic fluctuation level of the time-series.

	\subsection{Mean link-length analysis}		
			The second metric considered is the global mean link-length, $\langle d_{1n}\rangle$. If peaks often occur in a series (as in Fig. \ref{fig:example_dc_series}, panels (c) and (d)), points far from each other are not visible because far connections are hampered by peaks, and $\langle d_{1n}\rangle$ is consequently strongly reduced. The visibility of a generic node in the networks $G_{(c)}$ and $G_{(d)}$ is limited by the peak at $i=50$ (green-colored dot in Fig. \ref{fig:example_metr}), which in turn divides the networks into two main clusters (see bottom right panels of Fig. \ref{fig:example_metr}). The value of $\langle d_{1n}\rangle_{(c)}$ and $\langle d_{1n}\rangle_{(d)}$ are indeed much lower than $\langle d_{1n}\rangle_{(a)}$ and $\langle d_{1n}\rangle_{(b)}$, respectively, as can be seen in Fig. \ref{fig:example_metr}. On the other hand, $\langle d_{1n}\rangle$ is not essentially affected by the irregularities of a series. Indeed, irregularities mostly prevent medium-term connections than short and long-term links but, averaging over all nodes in the network, a value of the order of medium-term links is generally obtained for $\langle d_{1n}\rangle$. In fact, in Fig. \ref{fig:example_metr} the value of $\langle d_{1n}\rangle_{(b)}$ is approximately equal to $\langle d_{1n}\rangle_{(a)}$, while  $\langle d_{1n}\rangle_{(d)}$ is almost the same of $\langle d_{1n}\rangle_{(c)}$, indicating that there are no relevant changes in the global mean link-length due to irregularities. To conclude, the global mean link-length is strongly influenced by the occurrence of peaks, being slightly affected by the irregularities. The mean link-length measures how isolate and sporadic extreme events (i.e., peaks) are, with low $\langle d_{1n}\rangle$ values when the recurrence of peaks is high. Differently to high-order statistics (such as, for example, kurtosis), $\langle d_{1n}\rangle$ is able to fully capture the temporal dislocation of such extreme events along the time-series.
									
		\subsection{Combining $Tr$ and $\langle d_{1n}\rangle$}		
			As a consequence of the previous observations, the visibility algorithm turns out to be able to capture two main features of the temporal structure of a series: the recurrence of peaks and the presence of irregularities. It is worth noting that the two global (i.e., those associated to networks) measures analyzed so far inherit the local structural features of the mapped time-series. Therefore, a comparative temporal characterization of the time-series can be carried out by combining the behaviors of the global metrics.
			
			If $Tr$ and $\langle d_{1n}\rangle$ are focused on, a time-series can differ from another through a combination of the metrics behaviors, namely $Tr$ and $\langle d_{1n}\rangle$ can increase, decrease, or remain almost constant. Excluding the combination in which both $Tr$ and $\langle d_{1n}\rangle$ are almost constant (i.e., the two compared time-series share the same temporal features), four different \textit{cases} can occur and they are explained in table \ref{tab:modes}. Therefore, given the metric trends, it is possible to infer from table \ref{tab:modes} how time-series differ in terms of peaks and irregularities.

Being the degree centrality, $k$, a direct measure of the visibility of nodes, it can contemporarily account for both the recurrence of peaks and the presence of irregularities. In other words, $\langle k \rangle$ combines the features of both $Tr$ and $\langle d_{1n}\rangle$ in a single global metric. Therefore, due to its intrinsic definition, the degree variation in general cannot be univocally related to a specific temporal feature (either peaks or irregularities occurrence). For this reason, although being conceptually one of the easiest measure to interpret, here the degree centrality will be mainly discussed as a posteriori validation of the transitivity and global mean-length behaviors.


\section{Results}\label{sec:results}

The procedure described in the previous section is adopted to analyze the velocity time-series of the turbulent channel flow, starting from the streamwise component, $u^*$, and then considering the other velocity components, $v^*$ and $w^*$.
		
	\subsection{Streamwise velocity component, $u^*$}
		In Fig. \ref{fig:channel_metrics} the metrics $\left(\widetilde{k},\widetilde{Tr},\widetilde{d_{1n}}\right)$ are plotted as a function of the wall-normal coordinate, $y^+$. Substantial variations of these metrics occur moving along the wall-normal direction, exhibiting clear and regular trends. The three metrics have overall similar behaviors, rising from the wall up to a maximum value, then decreasing to $y^+\simeq 100-200$ and, finally, barely changing towards the center of the channel. The maximum values of $\widetilde{k}$, $\widetilde{Tr}$, and $\widetilde{d_{1n}}$ are not exactly at the same value of $y^+$, but they are quite close in the range $y^+\simeq 4-7$. The global network-metrics $\left\lbrace\langle k \rangle; Tr; \langle d_{1n}\rangle\right\rbrace$, computed for each of the $S_x \times S_z$ grid-point, have regular trends similar to the averaged ones shown in Fig. \ref{fig:channel_metrics}, which are thus representative of the global metrics measured along the wall-normal direction and in different $(x,z)$ coordinates (see also Fig. \ref{fig:hom_XZ} in Appendix C).
	
To infer the temporal structure of time-series along the wall-normal coordinate, we start from time-series close to the wall and then proceed towards the center of the channel. In particular, we focus on three representative $y^+$ stations, i.e. $y^+ = 0.017, 15.4, 996.3$.
 The three time-series, $u^*(t)$, at the selected $y^+$ stations are shown in Fig. \ref{fig:SerieU_x1601}(a), while a graphical representation of the corresponding networks is displayed in Fig. \ref{fig:SerieU_x1601}(b), revealing the presence of different topological features at different distances from the wall.

		Starting from the time-series at $y^+=0.017$, this series appears globally quite smooth, with relatively slow variations in time, resulting in few pronounced peaks. Moving from $y^+=0.017$ to $y^+=15.4$, the Fig. \ref{fig:channel_metrics} shows that the transitivity, $\widetilde{Tr}$, consistently increases, while the average mean link-length, $\widetilde{d_{1n}}$, and the average degree, $\widetilde{k}$, noticeably decrease. This combination of metrics corresponds to the \textit{Case D} in table \ref{tab:modes} (here $TS_{(1)}$ and $TS_{(2)}$ correspond to time-series at $y^+=15.4$ and $y^+=0.017$, respectively). A normalized time-series extracted at $y^+=15.4$ is then expected to be (on average) more regular than a series extracted at $y^+=0.017$ (indicated by the growth of $\widetilde{Tr}$), and with a more frequent occurrence of peaks (indicated by the drop of $\widetilde{d_{1n}}$). The reduction of $\widetilde{k}$ suggests that the increasing occurrence of peaks affects the global visibility more than the reduction in the irregularities. Looking at the time-series extracted at $y^+=15.4$ of Fig. \ref{fig:SerieU_x1601}(a), it is indeed with more peaks than the series extracted at $y^+=0.017$. This aspect is also evident in a more clustered topology of the network built on the time-series at $y^+=15.4$ (see Fig. \ref{fig:SerieU_x1601}(b)). The regularities appear globally similar but, as zoomed in the inset of Fig. \ref{fig:SerieU_x1601}(a), the two time-series are locally different. In particular, the time-series at $y^+=0.017$ appears more irregular, as indicated by the transitivity.
		
		From $y^+=15.4$ to $y^+=996.3$ (i.e., close to the center of the channel, $h$), all the average metrics substantially decrease. This combination of metrics corresponds to the \textit{Case C} in table \ref{tab:modes}. Accordingly, we expect that a time-series extracted at the center of the channel is (on average) less regular than a series at $y^+=15.4$ and with a more frequent recurrence of peaks. This behavior can be clearly seen in Fig. \ref{fig:SerieU_x1601} where the time-series at the center of the channel is more fluctuating than the time-series at $y^+=15.4$, and the corresponding network appears more clustered and disordered. It is interesting to note that from $y^+ \approx 10^2$ to the center of the channel, the three metrics barely change.

		In summary, the temporal features of the series are actually as predicted by combining the network metrics. Through the behavior of the metrics along the $y^+$ direction, Fig. \ref{fig:channel_metrics} yields first important results on the presence, dislocation and structure of extreme events and irregularities of the time-series. This kind of information can enrich the comprehension of the flow dynamics. It is important to remark that the behavior of a single metric is not a sufficient information, but a combination of the two metrics, $\widetilde{Tr}$ and $ \widetilde{d_{1n}}$, instead, determines how two time-series differ in terms of recurrence of peaks and/or irregularities. Moreover, we do not refer to the specific value assumed by the metric, but the analysis is comparative as it focuses on the trend each metric assumes as a function of the distance from the wall. Specifically, comparing time-series at the wall and at the center of the channel, peaks are expected to be remarkably closer, while irregularities do not substantially change ($\widetilde{d_{1n}}$ decreases while $\widetilde{Tr}$ slightly increases). In fact, as shown in Fig. \ref{fig:SerieU_x1601}(a), in the center of the channel peaks occur more frequently but the irregularity between them remains basically unvaried. However, this trend is not monotonic along $y^+$, since the time-series locally (around $y^+=15.4$) change their regularity. In terms of the network topology, as displayed in Fig. \ref{fig:SerieU_x1601}(b), close to the wall the network is composed by different subnetworks, corresponding to the peaks of the series, which are widely connected with each other and internally. Going towards $y^+=15.4$, the simultaneous decrease of $\widetilde{d_{1n}}$ and increase of $\widetilde{Tr}$ mainly break down long connections among the subnetworks. The drop of $\widetilde{d_{1n}}$ plays a major role here, acting to split long-term links. The subsequent decrease of both $\widetilde{d_{1n}}$ and $\widetilde{Tr}$ (from $y^+=15.4$ to $y^+=996.3$) breaks principally intra-network connections. At this stage, the prevailing effect is locally induced by the increase of irregularity, which leads to a ramification of each subnetwork.

\noindent A comment on the degree centrality can be eventually carried out. A high value of $\widetilde{k}$ indicates a globally convex time-series, while low values indicate a strong fragmentation of the visibility network \cite{donner2012}. As a result, considering the behavior of $\widetilde{k}$ in Fig. \ref{fig:channel_metrics}, at high values of $y^+$ the time-series are globally more fragmented than the time-series close to the wall, confirming what found observing the trends of $\widetilde{Tr}$ and $\widetilde{d_{1n}}$.

		Finally, the average cumulative degree distributions, $\widetilde{P_k}$, are illustrated in Fig. \ref{fig:Pk_cum_sety} (semi-log plot). As for the metrics, a degree distribution was computed for each network and all the distributions were then averaged over the homogeneous directions. The $P_k$ of a visibility network can be thought of as a measure of the (linear and nonlinear) temporal dependences existing in the time-series \cite{Luque2009}. However, differently from the horizontal visibility algorithm, the behavior of the degree distribution also depends on the probability density function (pdf) of the mapped time-series when the natural visibility algorithm is applied \cite{manshour2015_fractional}. As evident from the Fig. \ref{fig:Pk_cum_sety}, the tail of the distributions reveals decreasing exponential trends, i.e. the higher degree values (i.e., the \textit{hubs}) are generally very infrequent. In particular, the exponent of the fitting, $\gamma$, of the $\widetilde{P_k}$ increases (in modulus) from the wall towards the center of the channel, $y=h$. This is consistent with the temporal integral scale measurements \cite{Quadrio_Luchini}, which decrease from the wall to the center of the channel.

\noindent In order to isolate (from the pdf contribute) the net impact of (linear and nonlinear) dependences in the turbulent time-series, we built four series by shuffling four velocity time-series (at arbitrary $(x,z)$ locations) at the same wall-normal distances considered, i.e., $y^+=(0.017, 15.4, 106.2, 996.3)$. As shown in Fig. \ref{fig:Pk_cum_sety} (and highlighted in the inset), the slopes of $P_k$ from the shuffled series are substantially steeper than the turbulent time-series. This demonstrates the key role of the (linear and nonlinear) correlation aspects of the turbulent series.
			
		Up to now, we pointed out the ability of the visibility-based networks to shed light on the temporal structure of the corresponding mapped time-series. Now we try to relate the network metrics with the flow dynamics, that are responsible for the time-series behavior. Looking at the Fig. \ref{fig:channel_metrics}, three regions are particularly interesting (i.e. $y^+\lesssim7$, $7\lesssim y^+\lesssim150$, and $y^+\gtrsim150$) where the average metrics mostly change their trend. It should be noted that the values of $y^+$ delimiting such regions are very close to the limit values, $y^+=5$ and $y/h=0.1$, of the \textit{viscous sub-layer} and \textit{inner layer}, respectively \cite{pope}. In particular, for $Re_{\tau}=1000$, the inner layer limit is about $y^+=100$. The region for $y^+<5$ is characterized by slow moving fluid and the flow dynamics are dominated by the viscous shear stresses. The normalized time-series $u^*(t_i)$ here can be assumed to roughly share a similar temporal structure (although their mean and standard deviation values clearly change along $y^+$). The corresponding metrics (see Fig. \ref{fig:channel_metrics}) highlight this behavior resulting in barely increasing trends. As previously observed, around $y^+\simeq 4-7$ (which is the upper bound of the viscous sub-layer) the three metrics reach their maximum values. Here we expect, in terms of time-series shape, a minimum number of peaks along with the minimum irregularities. Recalling that all signals are normalized with the local mean and standard deviation, a possible interpretation is the following. Around $y^+\simeq 4-7$, we are approaching the \textit{buffer layer} ($5<y^+<30$), an intermediate region where viscous shear stress starts decreasing while turbulence activity begins to grow. However, at the very beginning ($y^+\simeq 4-7$), turbulent processes are very low, thus resulting in a minimum of irregularities, which act over a signal that is still affected by slow temporal variations (i.e., low number of peaks). The combination of these dynamics reasonably explains the maxima reached by all the metrics around the region $y^+\simeq 4-7$. For $y^+>5$ the flow dynamics are more affected by the Reynolds shear stresses, and the flow shows a tendency to organize into coherent turbulent patterns \cite{pope}. The structure of the time-series is then affected by turbulent processes (e.g., \textit{ejections} and \textit{sweeps} \cite{pope}), leading to rapid temporal variations. This behavior could be recognized in the drop of the average metrics (Fig. \ref{fig:channel_metrics}). As $y^+$ further increases ($y^+>100$), the turbulent patterns are less affected by the wall and they can develop in larger structures. However, the coexistence of multiple scales and the more complicated flow structure \cite{Jimenez2013} seems not to translate into a clear trend for the network metrics.
		
	\subsection{Transversal and spanwise velocity components, $(v^*,w^*)$}
		
		In Fig. \ref{fig:uvw}, the three metrics $\widetilde{Tr}$, $\widetilde{d_{1n}}$ and $\widetilde{k}$ are displayed for all the velocity components, $(u^*,v^*,w^*)$ to facilitate the comparison. In general, the mean link-length and the average degree measured on the time-series of $v^*$ and $w^*$ show trends similar to those of $u^*$, while different trends are obtained considering the transitivity.
		
More in particular, for $\widetilde{d_{1n}}$ the trends over $y^+$ for the three velocity components are similar, but values for the streamwise velocity, $u^*$, are overall higher than those displayed by $v^*$ and $w^*$. The relative difference decreases towards the center of the channel. The scenario for the degree centrality, $\widetilde{k}$, is analogous. Differences for the $\widetilde{k}$ values of the three components are marked close to the wall, while $\widetilde{k}$ values tend to coincide approaching the channel center. This behavior can be explained by considering that close to the wall the presence of the wall itself strongly influences and differently characterizes the flow dynamics in the three directions of the velocity, and consequently the networks based on the corresponding time-series are affected. On the contrary, the wall effects decrease moving far from the wall ($y^+>100$), thus differences among the metrics built on $u^*$, $v^*$ and $w^*$, reduce. As for the transitivity, $\widetilde{Tr}$, the metric difference among velocity components is even more accentuated. In fact, in the region $y^+<100$, not only values are different but also metrics display different trends. In particular, the wall-normal velocity component, $v^*$, is strongly affected by the presence of the wall (recall that close to the wall the motion corresponds to flow in planes parallel to the wall \cite{pope}) and this in turn involves the transitivity. For example, spikes with large negative values can be found in the time-series of $v^*$ as a consequence of strong \textit{events} that appear only in the very near-wall region, revealed by high kurtosis levels \cite{xu1996}. Since these deep peaks are negative and relatively short, the degree and the mean link-length of the corresponding networks are barely affected, while the transitivity is strongly reduced. Towards the channel center, similarly to the mean link-length $\widetilde{d_{1n}}$, the transitivity differences for the three velocity components tend to reduce.

In the end, the cumulative degree distributions, $\widetilde{P_k}$, of the networks built on the three velocity components, $(u^*,v^*,w^*)$, and averaged over the grid-points in the homogeneous directions are displayed in Fig. \ref{fig:Pk_cum_sety_UVW}. At fixed positions from the wall ((a): $y^+=0.0017$, (b): $y^+=15.4$, (c): $y^+=996.3$), the slope of the three components is pretty similar, confirming that a steeper decay is present when moving far from the wall (from $y^+=0.0017$ to $y^+=996.3$).

\section{Conclusions}\label{sec:concl}

		In this work, the application of the natural visibility graph to time-series of a fully-developed turbulent channel flow was studied. Our attention was focused on the streamwise velocity component, $u$, although the other velocity components were also explored. Velocity time-series were adopted to build the corresponding networks as the velocity field is one of the most intuitive quantity to characterize a fluid flow. However, the visibility graph method can be applied to other quantities of turbulence interest, such as the Reynolds shear stress, the kinetic energy, or the vorticity field. Firstly, we provided some novel insights into how the network metrics are affected by the different temporal structure of the mapped time-series. The average transitivity, $\widetilde{Tr}$, the here introduced mean link-length, $\widetilde{d_{1n}}$, and the average degree, $\widetilde{k}$, were chosen as the most representative metrics. Their trends turned out to effectively highlight the temporal features, in terms of peaks and irregularities, of the mapped time-series along the wall-normal coordinate, $y^+$. Furthermore, the cumulative degree distributions are found to show a decreasing exponential tendency, but with fitting exponent values at least one order of magnitude greater than uncorrelated random series. Different metrics variations were also quite well associated to the flow dynamics, as responsible of the time-series behavior.
		
		Despite several statistical techniques are available to study nonlinear time-series, specifically regarding turbulence, most of them are invariant under different temporal structures of the time-series. The visibility-network analysis, instead, reveled to be a powerful and synthetic tool to handle \textit{big-data} and to explore specific temporal features of the mapped series, without losing information about their temporal structures and also capturing the underlying flow dynamics. In fact, each network is built holding the temporal dislocation of important temporal features, such as extreme events and irregularities. To extract and handle this information is crucial for a deeper understanding of the flow dynamics, since the most common statistical tools adopted in turbulence, from spectral analysis to higher-order moments, are not able to retain the temporal collocation of such phenomena. Our network-based approach demonstrates that visibility graph method is able to give much information about temporal structure of turbulent time-series and it will deserve future efforts, such as community and neighborhood detection, to better explore the network topology and its physical meaning. Future works can also involve simulations with different Reynolds numbers, and weighted or directed networks may be considered. Furthermore, finer spatial and temporal simulation resolutions may be considered.
		
		Based on present findings, the proposed procedure may thus provide a promising support to the classical methods for accurate time-series analyses of inhomogeneous turbulent flows. In particular, given a time-series and the behaviors of the network metrics as a function of the distance from the wall, it is possible to qualitatively infer the behavior of the time-series at another wall-normal distance. This method can be then particularly useful as a predictive and supportive tool when experimental measurements are difficult.
				
\section*{Acknowledgments}
\addcontentsline{toc}{section}{\numberline{}Acknowledgments}
This work was supported by the scholarship \textit{Ernesto e Ben Omega Petrazzini}, awarded by the \textit{Accademia delle Scienze di Torino}, Turin (Italy). A special thank goes to the \textit{Accademia delle Scienze} and the Petrazzini family. The authors would also like to thank J. G. M. (Hans) Kuerten for the fruitful discussion of the results.

\appendix
\section*{Appendix A}

	The selected grid-points of the sub-domain $(S_x,S_y,S_z)\subset(N_x \times N_y \times N_z)$ are reported below (according to the labeling of the online database) in the form $(a:d:b)$, where $a$ and $b$ are respectively the first and the last index of a uniformly spaced interval, and $d$ is a grid step size (e.g., $(1:2:9)$ takes the indices $\left\lbrace 1;3;5;7;9\right\rbrace$):
\begin{itemize}
	\item wall-normal direction, \[ Y = \left \{
	\begin{array}{l}
	(0:1:21)\\
	(23:2:39) \\
	(42:3:54) $ and $ (58:3:79) \\
	(84:5:169)\\
	(179:10:239) \\
	255$, i.e. $y^+=996.3$;$
	\end{array}
		\right.
	\]
	\item streamwise direction, $X=(0:32:2016)$;
	\item spanwise direction, $Z=(110:128:1518)$.
\end{itemize}

\section*{Appendix B}

A sensitivity analysis on the number of nodes $N$ is here reported, by varying the temporal discretization and consequently the cardinality of the corresponding network. We recall that $\delta t=0.0065$ leads to a number of nodes, $N=4000$. Two other time steps, namely $2 \, \delta t$, and $4 \, \delta t$, have been considered, resulting in networks with $N=2000$ and $N=1000$, respectively. In Fig. \ref{fig:decamp}, the metrics as function of $y^+$ are displayed for the three temporal samplings, $c\, \delta t$, with $c=1, 2, 4$. Mean link-length and degree centrality are reported as scaled with $c$ (namely $c\, \widetilde{d_{1n}}$ and $c\, \widetilde{k}$), to facilitate the comparison between samplings. The transitivity, $\widetilde{Tr}$, is not scaled with $c$ as by definition varies between 0 and 1. It can be observed that, apart from the specific values reached by the transitivity, the metrics behavior along the wall-normal direction $y^+$ is not sensitive to the choice of the temporal discretization (i.e., the number of nodes).

\section*{Appendix C}

In this section we report few representative plots of the global metrics calculated on networks at different single streamwise, $x$, and spanwise, $z$, locations. In Fig. \ref{fig:hom_XZ} we illustrate the behavior of the transitivity, the mean link-length and the degree centrality as a function of $y^+$ before the averaging operation, performed according to the Eq. (\ref{eq:aver_measures}), and compare them with the averaged behavior (as shown in Fig. \ref{fig:channel_metrics}). Specifically, in Fig. \ref{fig:hom_XZ}, we plotted $48$ curves for each metric, obtained from $48$ uniformly spaced grid-points in the $(x,z)$ directions and covering the whole domain. As can be seen, the averaged behaviors (reported in black in Fig. \ref{fig:hom_XZ}) are representative of the behavior of the global metrics for different streamwise and spanwise grid-points. The distributions of the mean link-length and the degree centrality appear less noisy (especially at the center of the channel) than the plots of the transitivity because the latter is globally evaluated for each network (see Eq. (\ref{eq:Trans})), while $\left\langle k \right\rangle$ and $\langle d_{1n}\rangle$ are defined as averages over nodes (see Eq. (\ref{eq:DC_m}) and Eq. (\ref{eq:d1n_m})). Therefore, we can conclude that the statistics homogeneity of the flow is inherited by the visibility networks, making the average behavior along $y^+$ statistically meaningful.

\newpage

\section*{Figure Legends}

\noindent \textbf{Fig. 1} \textit{(Left)} Examples of sine time-series with different temporal features. In panels (a) and (b), the green-colored dot indicates the point $s(t_i=34)$, while yellow points highlight its neighborhood, $\Gamma_{34}$. In panels (c) and (d), the green-colored dot evidences the point $s(t_i=50)$. \textit{(Right)} Networks corresponding to the time-series on the left.

\begin{figure}[h]
			\flushright
			\includegraphics[width=.8\linewidth]{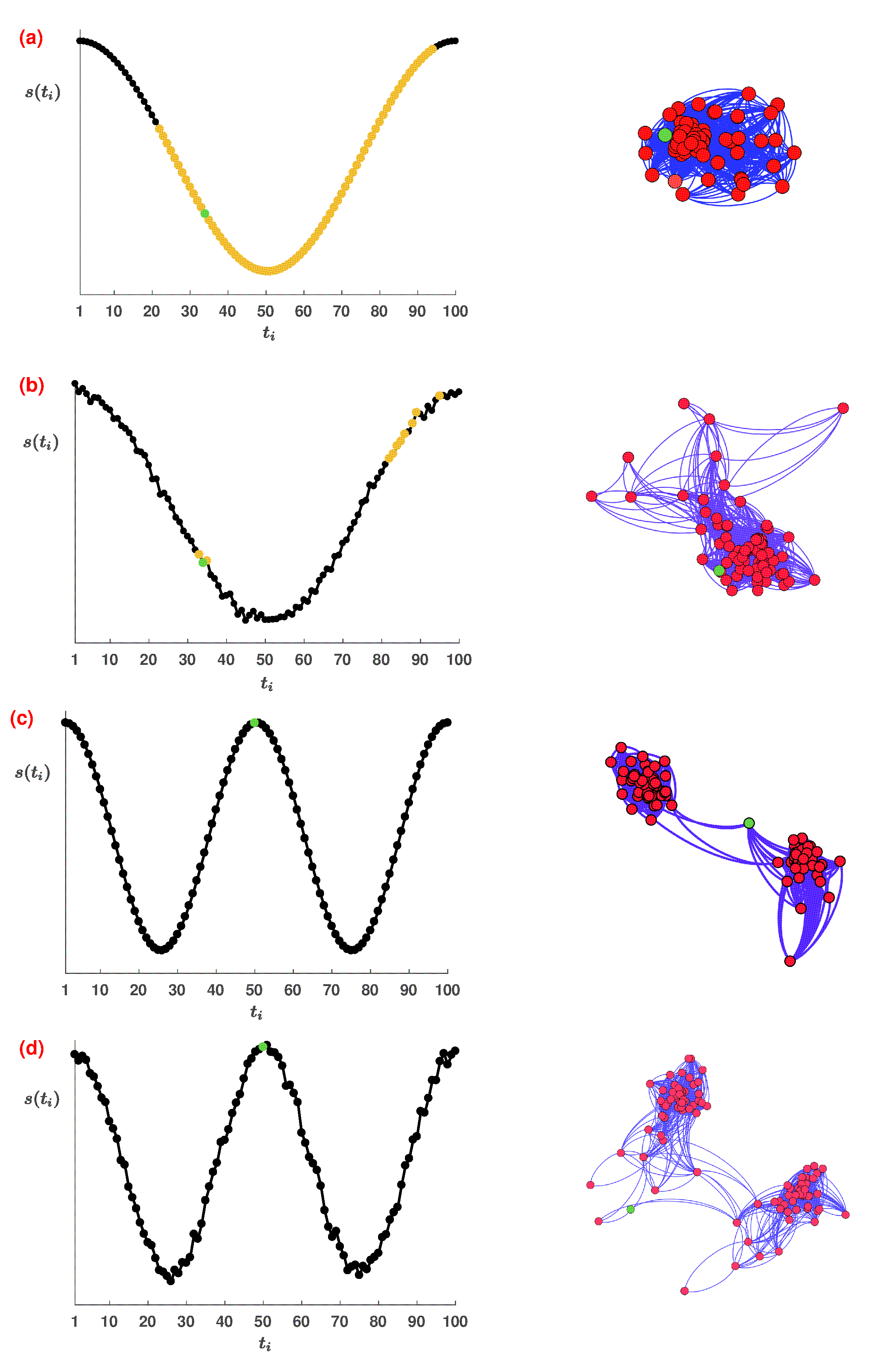}
			\caption{\label{fig:example_dc_series}}
		\end{figure}

\begin{figure}[h]
			\flushright
			\includegraphics[width=.6\linewidth]{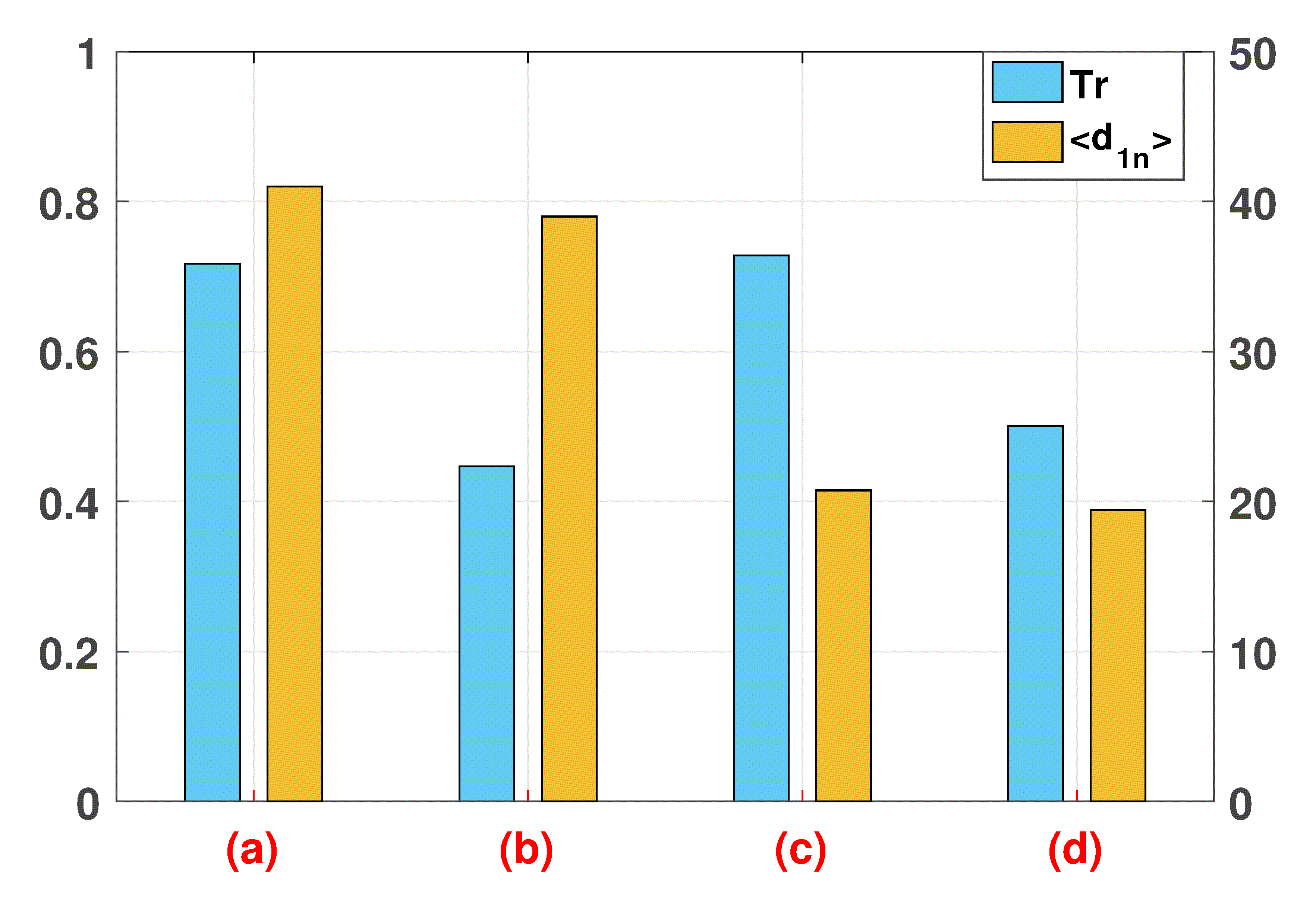}\hspace*{3.8cm}
			\caption{\label{fig:example_metr}}
		\end{figure}

\noindent \textbf{Fig. 2} Bar plot of the transitivity (blue) and global mean link-length (yellow) values for the four time-series of Fig. \ref{fig:example_dc_series}.

\noindent \textbf{Fig. 3} Global metrics averaged over the $S_x\times S_z$ networks as function of $y^+$, reported in a log-linear plot. The values of the metrics at the wall ($y^+=0$) are $\left(\widetilde{k},\widetilde{Tr},\widetilde{d_{1n}}\right)=\left(1.9995,0,1 \right)$, resulting from constant time-series and thus not shown here. Three representative values of $y^+$ are also highlighted.

\noindent \textbf{Fig. 4} \textit{(a)} Normalized time-series, $u^*(t_i)$, at the grid-points $X=1601$, $Z=750$ and $y^+= 0.017, 15.4, 996.3$. The choice of the coordinate in the homogeneous directions, $x$ and $z$, is arbitrary. In the inset the time-series at $y^+= 0.017$ (blue) and $y^+= 15.4$ (red) are highlighted and compared in the range $t_i\in[300,800]$. \textit{(b)} Graphical representation of the networks extracted from the time-series of panel \textit{(a)}.

\noindent \textbf{Fig. 5} Cumulative degree distributions, $\widetilde{P_k}$, averaged over the grid-points in the homogeneous $x$ and $z$ directions, reported in a linear-log plot. The distributions close to the vertical axis (highlighted in the inset) correspond to networks built on shuffled time-series at wall-normal distances: $(\times)$, $y^+=0.017$; $(\circ)$, $y^+=15.4$; $(\square)$, $y^+=106.2$; $(\vartriangle)$, $y^+=996.3$. The fittings were performed as $P_k\sim \exp (\gamma k)$, with a \textit{trust-region} method of optimization. The resulting values of $\gamma$ are $(-1.01,-1.08,-2.85,-2.86)\cdot 10^{-2}$ from $y^+=0.017$ to $y^+=996.3$, respectively; the slope for the shuffled series is about $\gamma\approx 2\cdot 10^{-1}$. The coefficient of determination, $R^2$, of the fittings is always above 0.99.

\noindent \textbf{Fig. 6} Average metrics $\widetilde{Tr}$, $\widetilde{d_{1n}}$, and $\widetilde{k}$ evaluated from time-series extracted from the velocity field, $(u^*,v^*,w^*)$.

\noindent \textbf{Fig. 7} Cumulative degree distributions of the networks built on the time-series of three velocity components, $(u^*,v^*,w^*)$, and averaged over the grid-points in the homogeneous directions. (a): $y^+=0.0017$, (b): $y^+=15.4$, (c): $y^+=996.3$.

\begin{figure*}[h]
		\flushright
		\includegraphics[width=\linewidth]{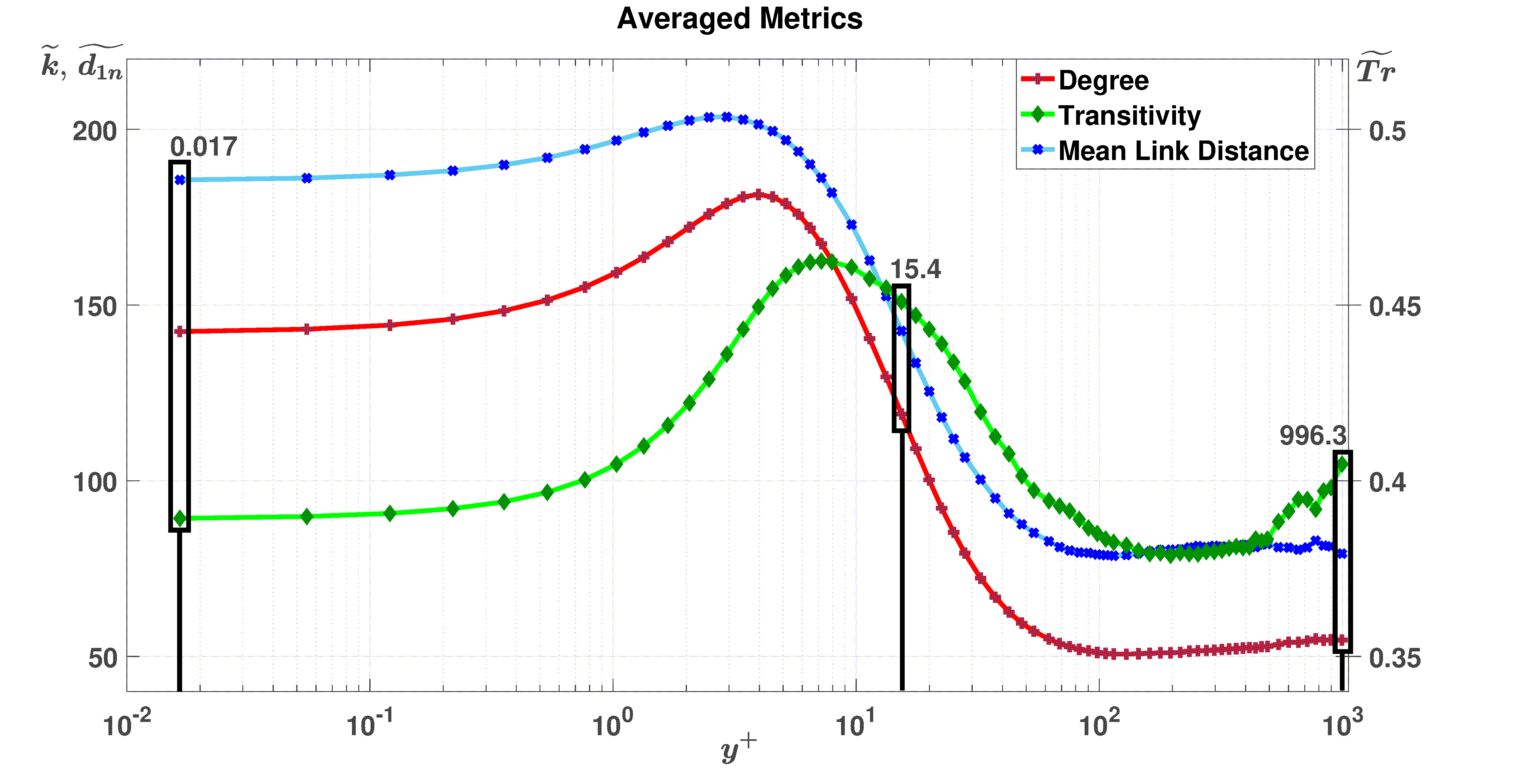}
		\caption{\label{fig:channel_metrics}}
	\end{figure*}

\begin{figure*}[h]
		\flushright
		 \includegraphics[width=\linewidth]{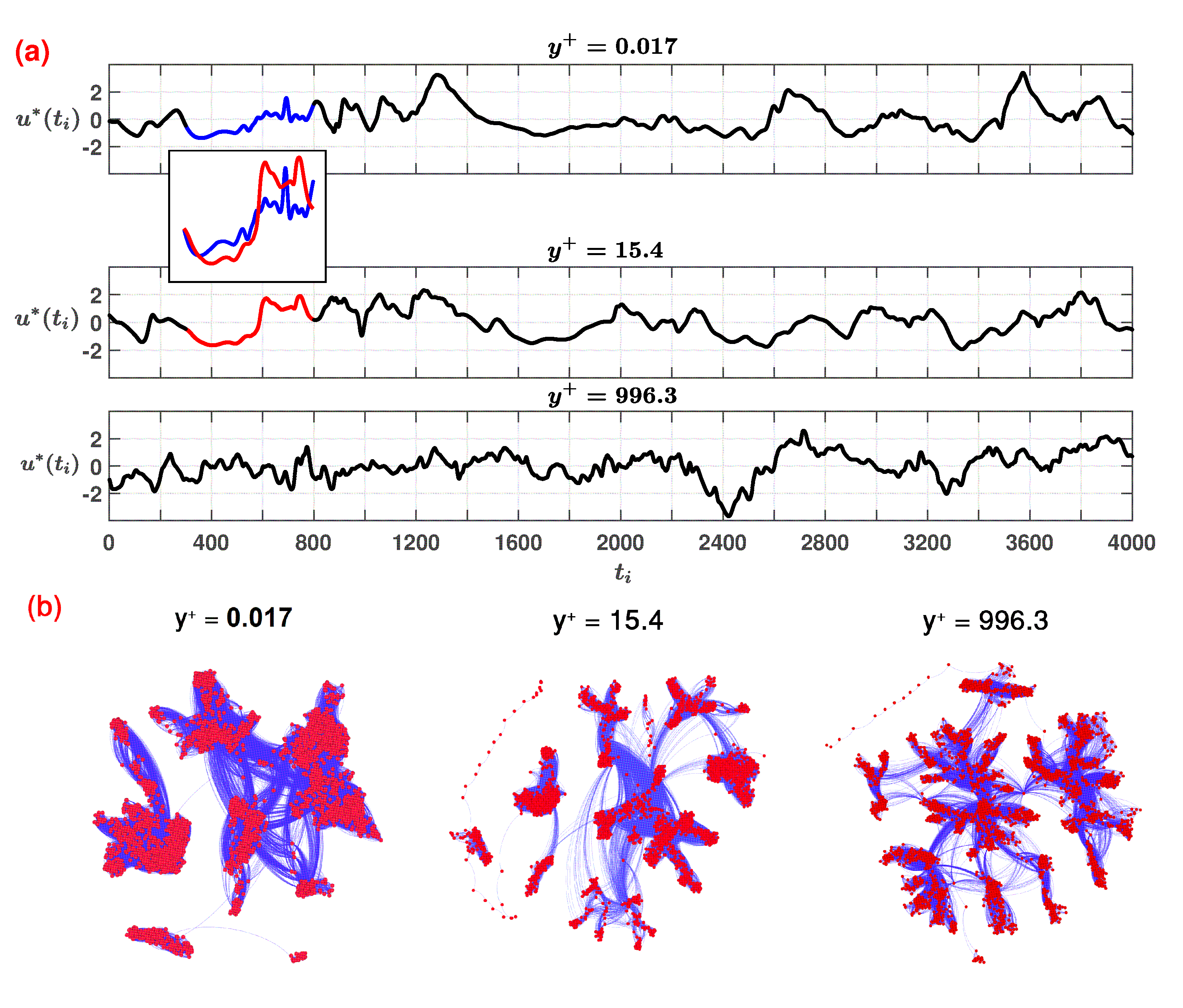}
		\caption{\label{fig:SerieU_x1601}}
	\end{figure*}

\begin{figure}[h]
			\flushright
			\includegraphics[width=.8\linewidth]{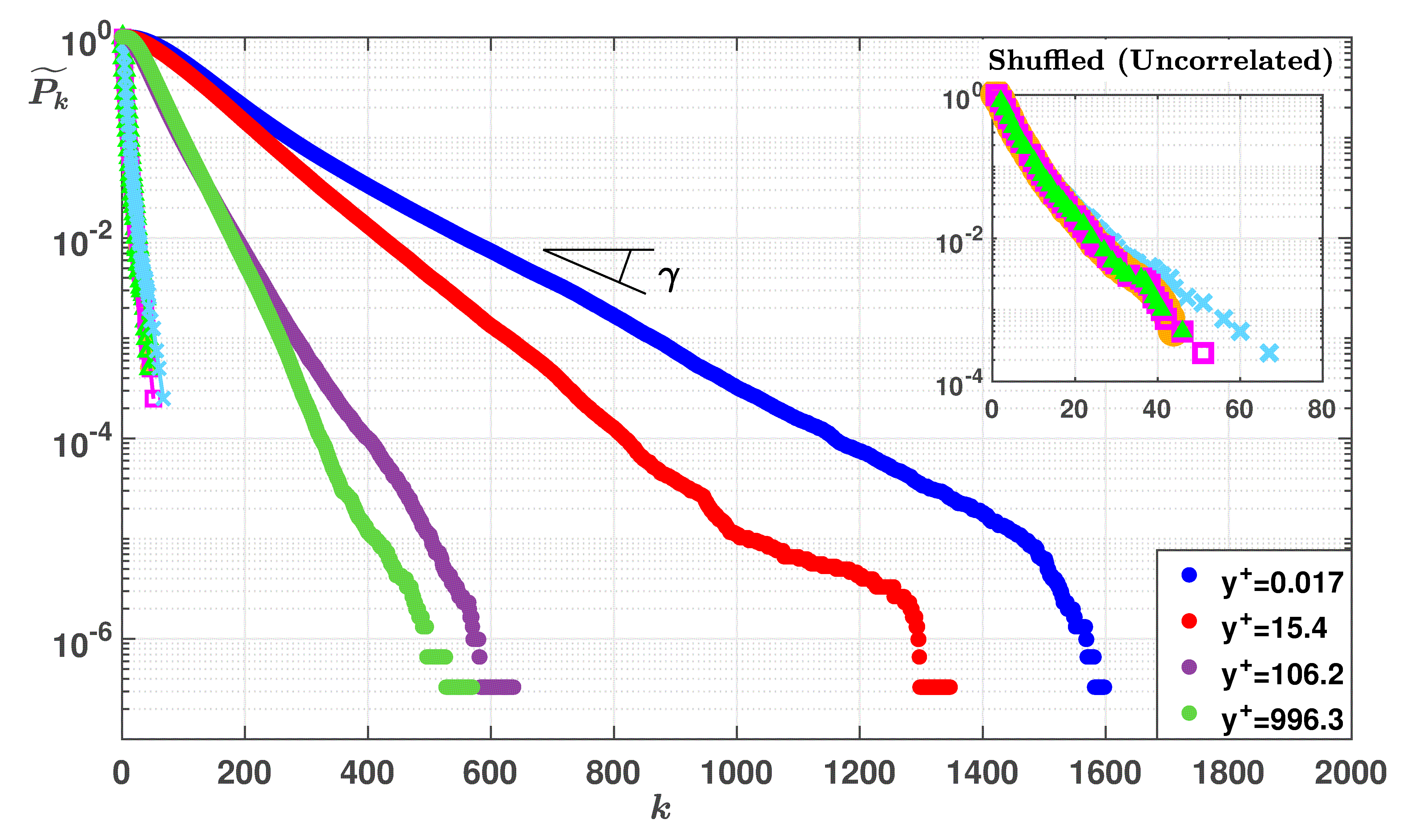}
			\caption{\label{fig:Pk_cum_sety}}
		\end{figure}

\begin{figure*}[h]
		\flushright
		\includegraphics[width=\linewidth]{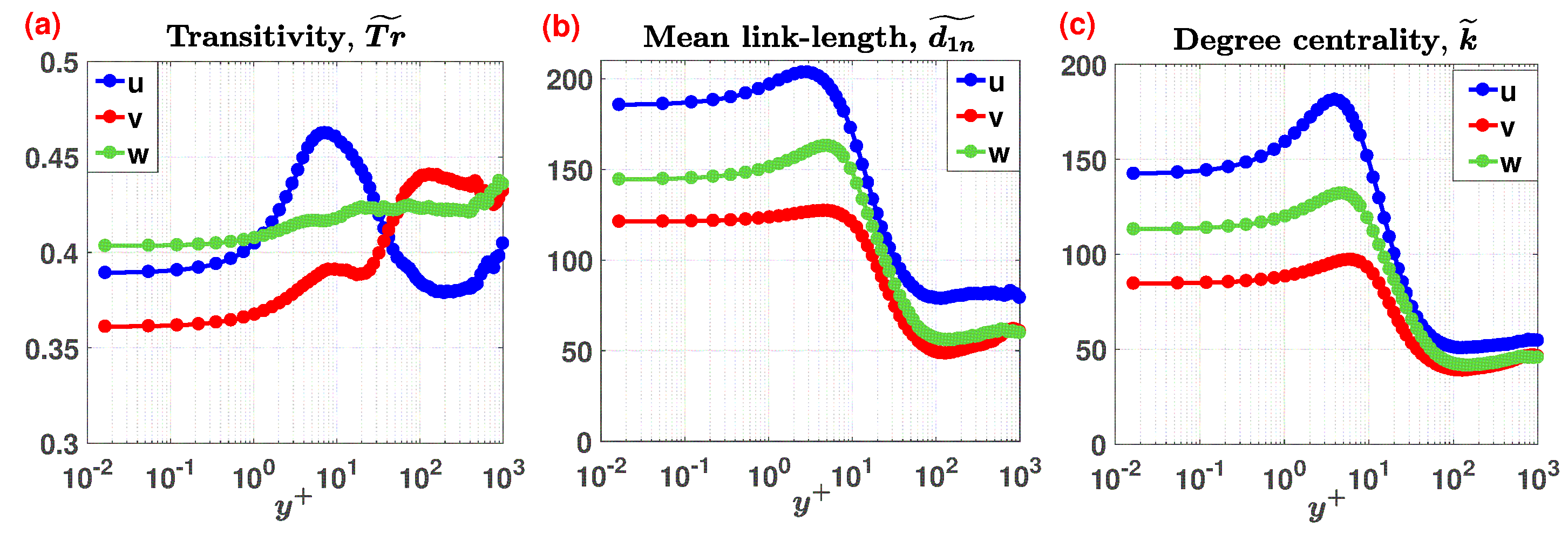}
		\caption{\label{fig:uvw}}
	\end{figure*}

\begin{figure}[h]
		\flushright
		\includegraphics[width=\linewidth]{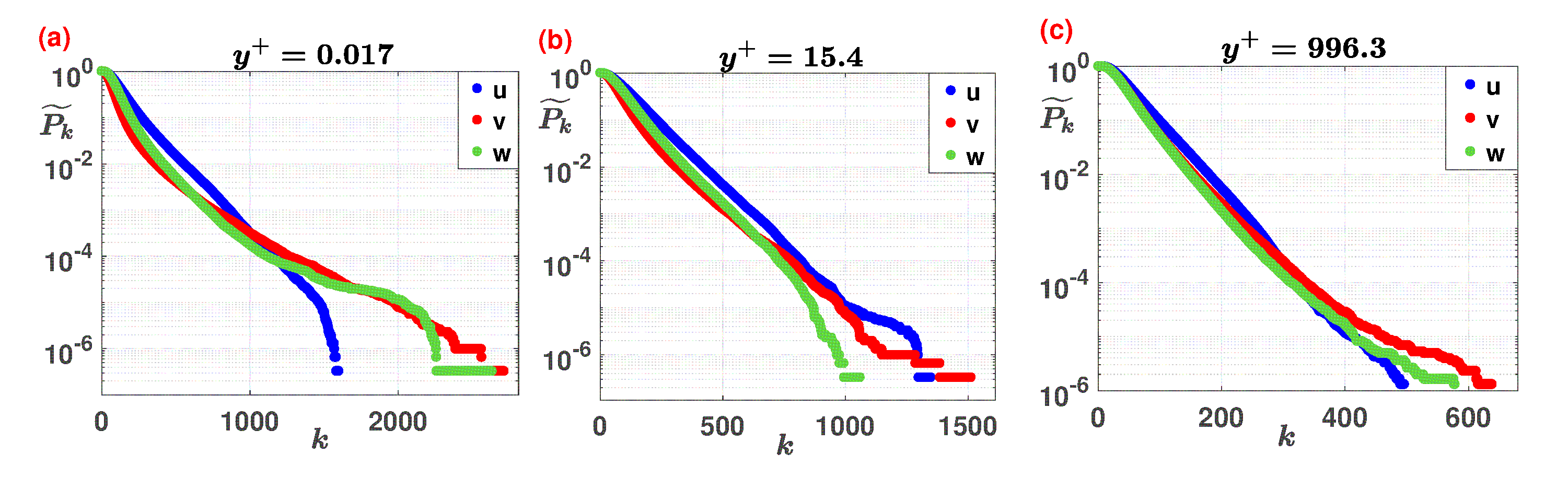}
		\caption{\label{fig:Pk_cum_sety_UVW}}
	\end{figure}

\begin{figure}[h]
	\renewcommand{\thefigure}{B\arabic{figure}}

		\flushright
		\includegraphics[width=\linewidth]{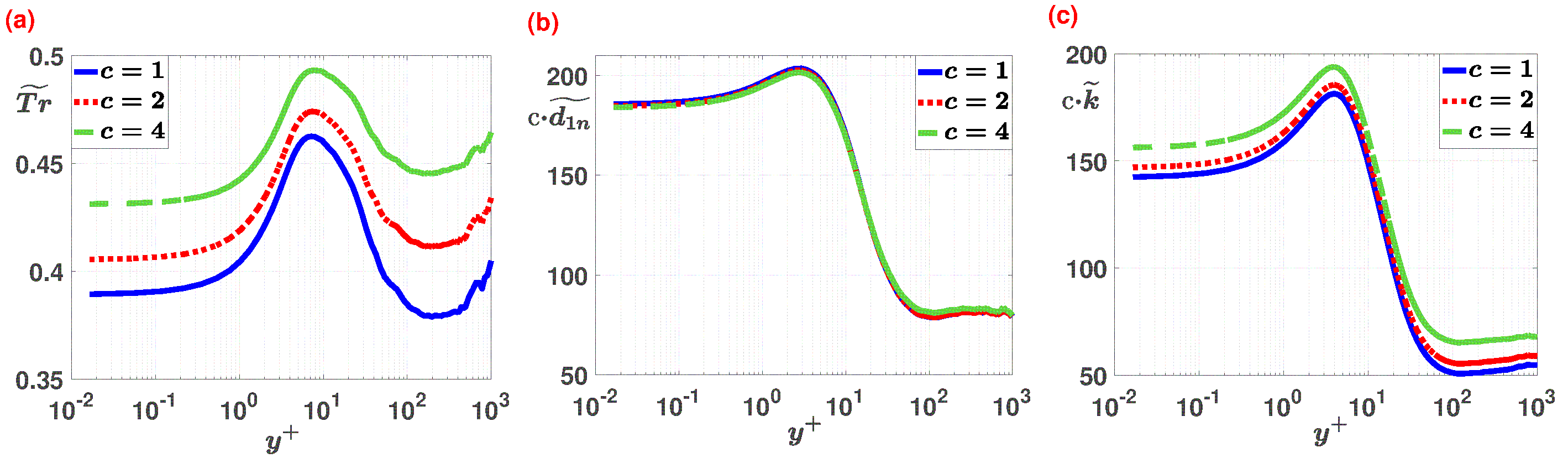}
		\caption{\label{fig:decamp}}
	\end{figure}

\begin{figure}[h]
	\renewcommand{\thefigure}{C\arabic{figure}}

		\flushright
		\includegraphics[width=\linewidth]{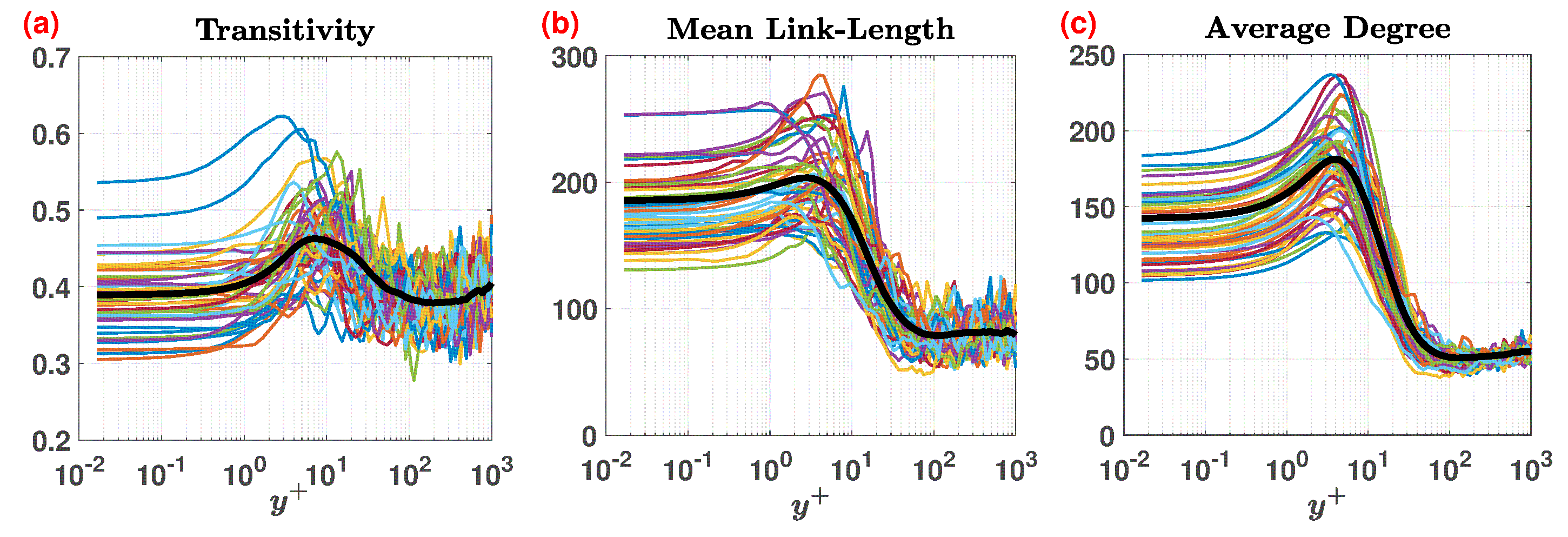}
		\caption{\label{fig:hom_XZ}}
	\end{figure}

\noindent \textbf{Fig. B1} Averaged metric behaviors, $\widetilde{Tr}$, $c \widetilde{d_{1n}}$, $c \widetilde{k}$, as function of $y^+$ for three different time sampling of the streamwise velocity time-series, $u^*$. The curves are obtained with $c \delta t$, where the sampling is $c=1$ (blue), which corresponds to the case in Fig. \ref{fig:channel_metrics}, $c=2$ (red), $c=4$ (green).

\noindent \textbf{Fig. C1} Metrics behaviors of networks built on the streamwise velocity component, $u^*$, and extracted at $48$ different uniformly spaced $(x,z)$ locations. The black plots correspond to the averaged behavior, as shown in Fig. \ref{fig:channel_metrics}.

\newpage

	\begin{table}[h]
		\flushright
		\caption{\label{tab:modes}Scheme of the ways two time-series, $TS_{(1)}$ and $TS_{(2)}$, can differ and corresponding behaviors of the global network-metrics, $Tr$ and $\langle d_{1n}\rangle$.}
		\footnotesize
		\renewcommand{\arraystretch}{1.8}
		
		\begin{tabular}{|c|c|c|c|c|}
			\hline
			\multicolumn{1}{|p{.1\linewidth}|}{\centering \normalsize \textbf{Cases}} & \multicolumn{1}{|p{.34\linewidth}|}{\centering \normalsize \textbf{Temporal structure features}} & \multicolumn{1}{|p{.25\linewidth}|}{\centering \normalsize \textbf{Metric behaviors}} & \multicolumn{1}{|p{.08\linewidth}|}{\centering \normalsize \textbf{TS1}} & \multicolumn{1}{|p{.08\linewidth}|}{\centering \normalsize \textbf{TS2}}\\
			\hline
			\textit{Case A} & \multicolumn{1}{|p{.34\linewidth}|}{\centering Peaks occur more frequently in $TS_{(2)}$ than in $TS_{(1)}$} & \multicolumn{1}{|p{.25\linewidth}|}{\centering $Tr_{(2)} \approx Tr_{(1)}$ $\langle d_{1n}\rangle_{(2)} < \langle d_{1n}\rangle_{(1)}$} & \raisebox{-5mm}{\includegraphics[width=.075\linewidth]{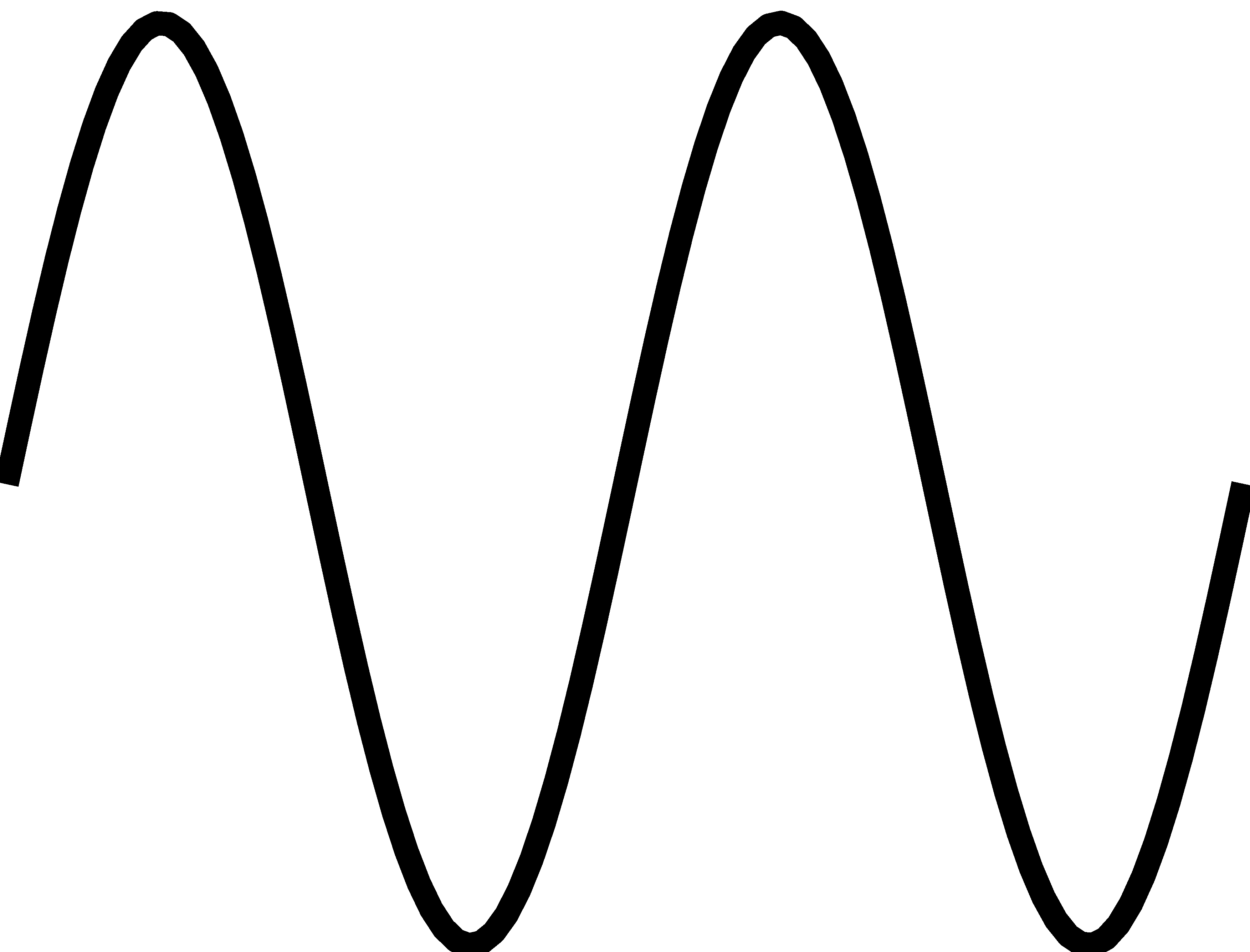}} & \raisebox{-5mm}{\includegraphics[width=.075\linewidth]{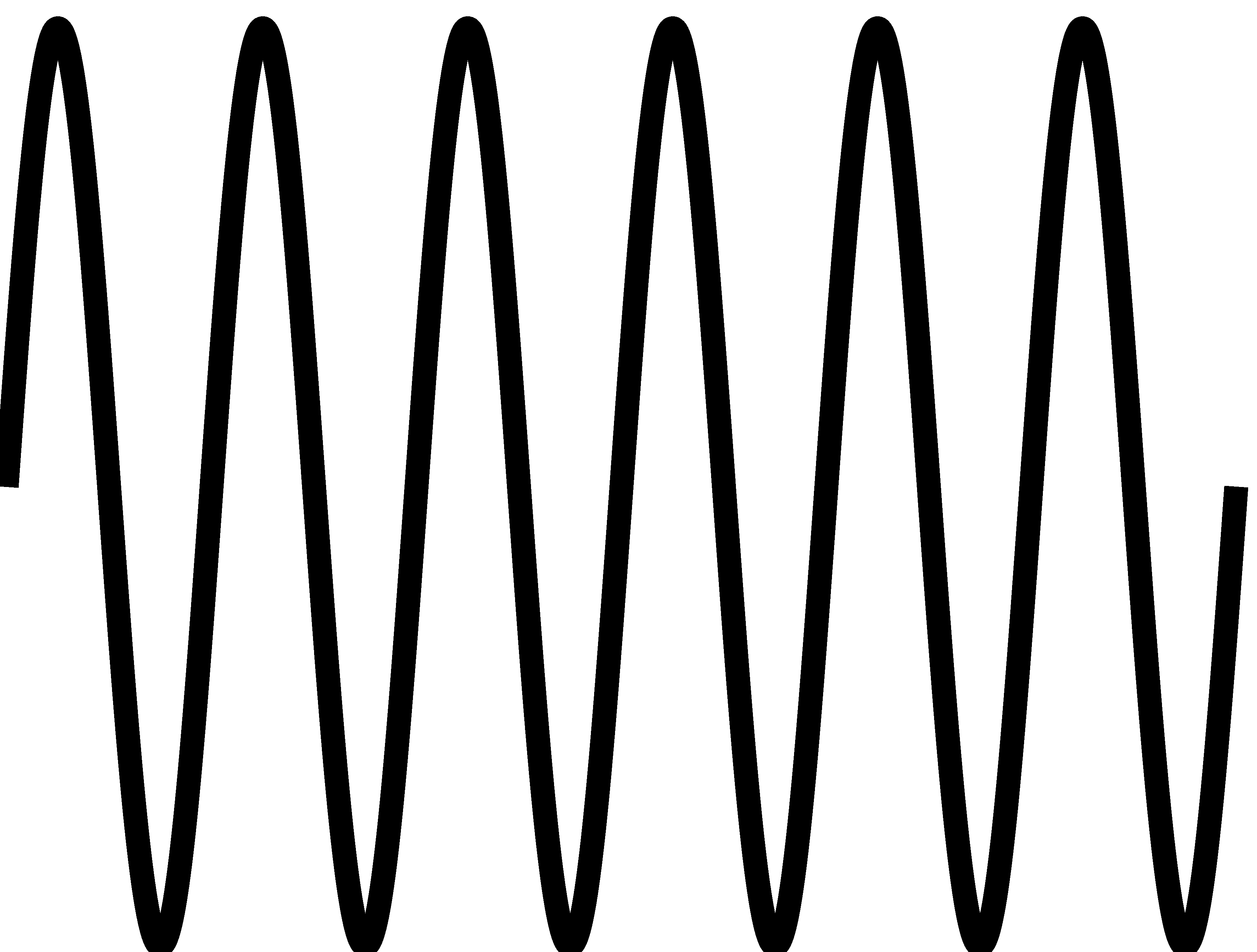}}\\
			\hline
			\textit{Case B} & \multicolumn{1}{|p{.34\linewidth}|}{\centering $TS_{(2)}$ is more irregular than $TS_{(1)}$} & \multicolumn{1}{|p{.25\linewidth}|}{\centering $Tr_{(2)} < Tr_{(1)}$ $\langle d_{1n}\rangle_{(2)} \approx \langle d_{1n}\rangle_{(1)}$} & \raisebox{-5mm}{\includegraphics[width=.075\linewidth]{Table1fig1.png}} & \raisebox{-5mm}{\includegraphics[width=.075\linewidth]{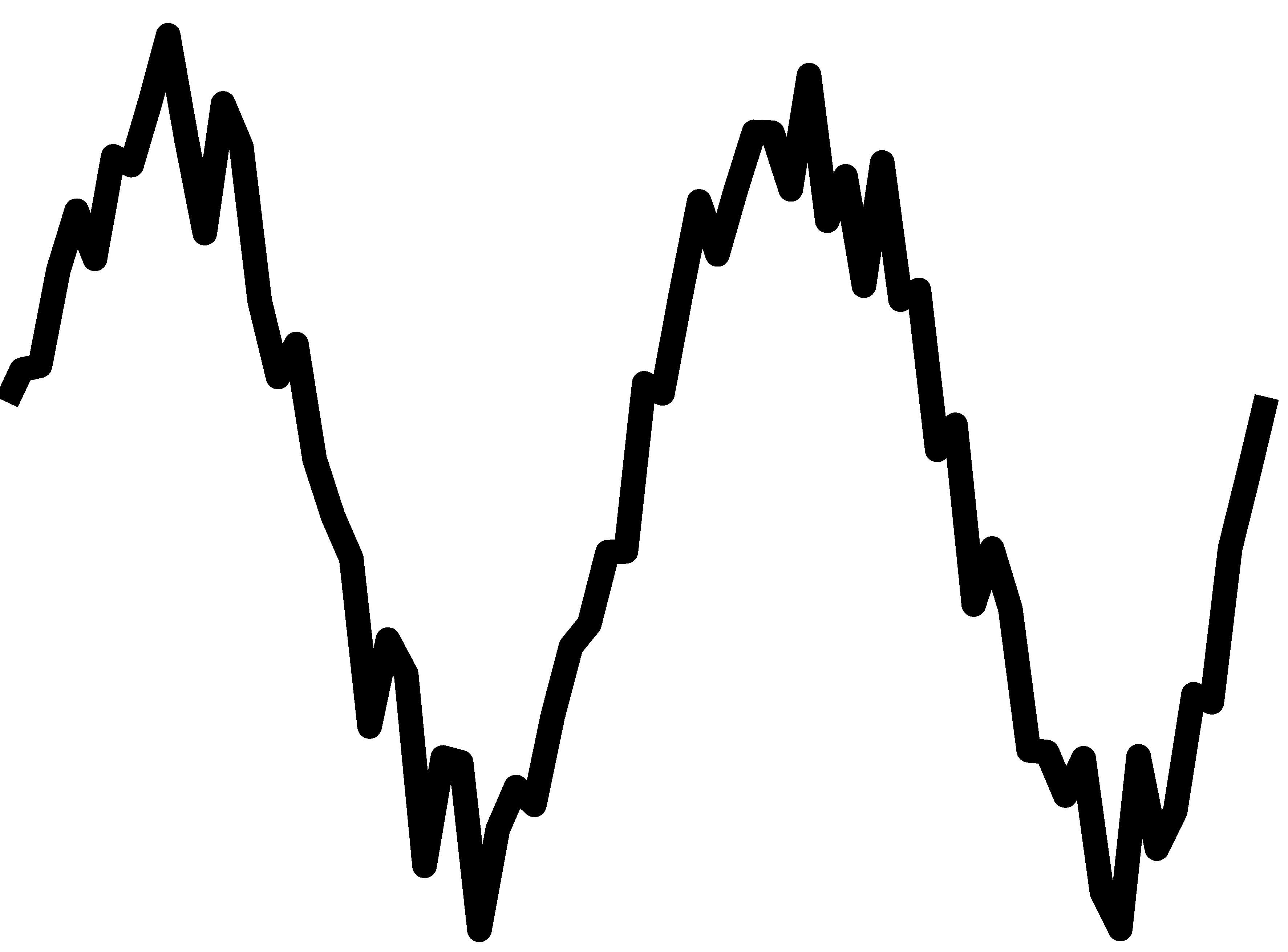}}\\
			\hline
			\textit{Case C} & \multicolumn{1}{|p{.34\linewidth}|}{\centering Peaks occur more frequently in $TS_{(2)}$ than in $TS_{(1)}$, and $TS_{(2)}$ is more irregular than $TS_{(1)}$} & \multicolumn{1}{|p{.25\linewidth}|}{\centering $Tr_{(2)} < Tr_{(1)}$ $\langle d_{1n}\rangle_{(2)} < \langle d_{1n}\rangle_{(1)}$} & \raisebox{-7mm}{\includegraphics[width=.075\linewidth]{Table1fig1.png}} & \raisebox{-7mm}{\includegraphics[width=.075\linewidth]{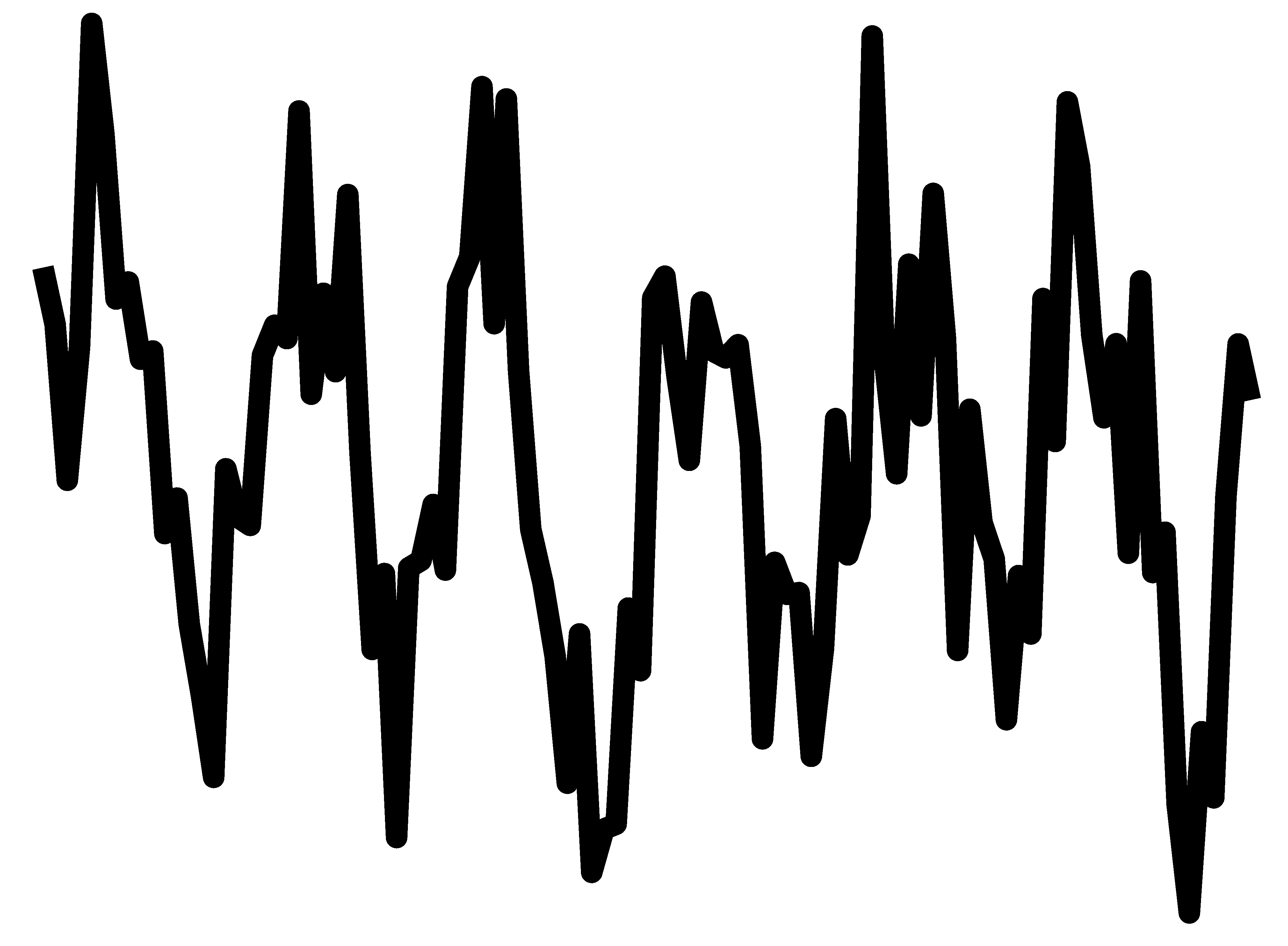}}\\
			\hline
			\textit{Case D} & \multicolumn{1}{|p{.34\linewidth}|}{\centering Peaks occur less frequently in $TS_{(2)}$ than in $TS_{(1)}$, and $TS_{(2)}$ is more irregular than $TS_{(1)}$} & \multicolumn{1}{|p{.25\linewidth}|}{\centering $Tr_{(2)} < Tr_{(1)}$ $\langle d_{1n}\rangle_{(2)} > \langle d_{1n}\rangle_{(1)}$} & \raisebox{-7mm}{\includegraphics[width=.075\linewidth]{Table1fig2.png}} & \raisebox{-7mm}{\includegraphics[width=.075\linewidth]{Table1fig3.png}}\\
			\hline	
		\end{tabular}	
	\end{table}

\end{document}